# Adiabatic Joule Heating of Copper from 4 K to the Melting Temperature.

Alain Guillet[1,a)] and Fabrice Delamarre[2]

[1]Groupe de Physique des Matériaux, U.M.R. CNRS 6634, Université and INSA of Rouen, Av. de l'Université, St Etienne du Rouvray, 76801 France.

[2]Département de Mécanique, INSA de Rouen, Av. de l'Université, St Etienne du Rouvray, 76801, France.[1]

**Abstract :**

Considering a copper wire heated by Joule effect and the variation of its resistivity and specific heat with temperature, we established numerical and analytical solutions (between 293 and 1356 K for the latter) for the evolution of its temperature over time. The Temperature vs. Time evolution follows a Lambertian function. The calculations are based on the assumption of adiabatic heating and uniform current distribution within the wire. We demonstrate that at very low temperature the heating rate is strongly dependent on copper purity.

When an electric current flows through a metallic conductor its temperature increases due to Joule effect, the knowledge of which is important in a number of scientific and technical fields. Consider, for examples, the copper matrix composites used for the winding of coils for the production of very high magnetic fields[1-4], the development of electromagnetic gunnery[5], the electrical explosion of wires[6,7] and the thermal ageing of solders in electronic components due to Joule heating[8]. Ohmic heating could also be an energy saving solution for metals heat treatment and a new route to investigate the influence of very high electrical heating rate on microstructures[9-16]. The temperature evolution of a conductor during electrical heating is an old problem[17]. It requires to solve numerically a differential equation which takes into account the energy losses due to convection, conduction and possibly radiancy[18-20]. However, if the current density is high and the annealing time short, adiabatic conditions are generally assumed, with the exception of Zielinskia et al.[21]. Therefore, the heating rate and the temperature increase are numerically

---

[a)] Author to whom correspondence should be addressed. Electronic address: Alain.Guillet@insa-rouen.fr

calculated by solving equation (1) which derives from an energy balance[22,23], but the analytical solution of equation (1) is lacking:

$$\int_0^{t_f} J^2 dt = \int_{T_0}^{T_f} \frac{D c_p(T)}{\rho(T)} dT. \tag{1}$$

Where, T is the temperature, t the time, J(t) the current density, D the material density, $C_p(T)$ the specific heat at constant pressure and $\rho(T)$ the electrical resistivity. $T_0$ and $T_f$ are the initial and final temperature, respectively. Solving equation (1) either numerically or analytically necessitates to take into account the temperature dependence of the materials properties D, $C_p$ and $\rho$. Surprisingly, some authors[13-16], consider D, $C_p$ and $\rho$ as constants, under these assumptions according to equation (1), the heating rate is given by (2):

$$\frac{dT}{dt} = \frac{\rho}{D C_P} J^2(t). \tag{2}$$

Then, following integration of (2) the temperature increase with time is given by equation (3):

$$T = T_0 + \frac{\rho}{D C_P} \int_0^t J^2(t') dt', \tag{3}$$

However, since $C_p$ and $\rho$ increase significantly between 4 K and the melting temperature of the metal, this is a gross approximation, although the temperature dependence of the density D, due to the thermal expansion, can be neglected. Indeed, according to the authors own calculations, not detailed here, the density variation of copper between 4 and 1356 K is only 6.5% even if the temperature dependence of the volumetric coefficient of thermal expansion is taken into account. We show below that $C_p$ and $\rho$ variations are indeed much greater than D variation. In this work we first calculate numerically the temperature dependence of the heating rate during Joule heating of copper by taking into account the temperature dependence of the materials properties $\rho$ and $C_P$. Then, we establish the heating time required to reach the melting temperature starting from three different initial temperatures : 4 K (boiling of liquid helium), 77K (boiling of liquid nitrogen) and 293K (ambient temperature). Finally, equation (1) is first solved numerically and then analytically for $T_0 > 293$ K. A constant current density of $J=10^9$ A/m$^2$ was chosen, typical, for example, of the load of a truck battery (I≈1000 A) applied to a 1 mm² cross-section wire. The copper properties required for the calculations are given in Table I.

TABLE I. Physical properties of copper.

| M[a] (kg/mole) | $\theta_D$[b,(24)] (K) | $T_m$[c] (K) | $\rho_{293K}$[c,(25)] ($10^{-8}\Omega$.m) | $\alpha$[d,(25)] (K$^{-1}$) | $C_{p(293K)}$[e] (J/kg/K) | D[f] (kg/m³) |
|---|---|---|---|---|---|---|
| 0.0635 | 315 | 1356 | 1.724 | 0.00393 | 393 | 8920 |

[a] is the atomic mass, [b] is the Debye temperature, [c] is the melting temperature,
[d] is the temperature coefficient of resistivity, [e] is the specific heat et T=293 K,
[f] is the density

For metals, both free electrons and phonons contribute to the specific heat, although the latter is largely preponderant at high temperature[27]. The phonons contribution to the specific heat at constant volume (in J/mole/K) is given by equation (4)[27]:

$$C_V(T) = 9R \left(\frac{T}{\theta_D}\right)^3 \left(\int_0^{\theta_D/T} \frac{x^4 e^x}{(e^x - 1)^2} dx\right). \quad (4)$$

There is no exact solution for this integral, which must be solved numerically. However, at high temperature the molar specific heat is close to $C_V \approx 3R$ (Dulong and Petit Law). At low temperature ($T/\theta_D \ll 1$) a T³ law is obeyed[27,28], then equation (5) follows :

$$C_v^{ph.}(J/mole/K) = 1450 \left(\frac{T}{\theta_D}\right)^3 \approx 4.65.10^{-5} T^3. \quad (5)$$

Furthermore, there is also an electronic contribution to the specific heat[26,27] which is only significant at low temperature (equation 6):

$$C_v^{el.}(J/mole/K) = 5.06 \times 10^{-4} T. \quad (6)$$

However, solving (1) requires the knowledge of $C_P$, the molar specific heat at constant pressure. At low temperature the difference between $C_p$ and $C_V$ is small[28] but it cannot be neglected *a priori* for these calculations. The relationship between $C_V$ and $C_P$ is given by (7)[26-28]:

$$C_P - C_V = \frac{9\alpha^2 VT}{\beta}. \quad (7)$$

with :

α: linear thermal expansion coefficient

V: molar volume

T: Temperature

β: Bulk modulus

According to Newham[26] the quantity A for copper (A=1.55x10$^{-5}$ mole/J) being constant over the temperature range 100-1200K, then equation (7) can be written as (8),:

$$C_P - C_V = ATC_P^2. \qquad (8)$$

Solving (8) for Cp, there is only one positive solution (9):

$$C_p = \frac{1-\sqrt{1-4ATC_v}}{2AT}. \qquad (9)$$

Finally, the total specific heat at constant pressure in J/kg/K is simply obtained by dividing (9) by the atomic mass M (in kg/mole) of copper (Table I). The evolution of $C_p$ and $C_v$ from 4K to the melting temperature of copper appears in Fig. 1. Clearly, $C_P$ becomes significantly different from $C_V$ only above 300K.

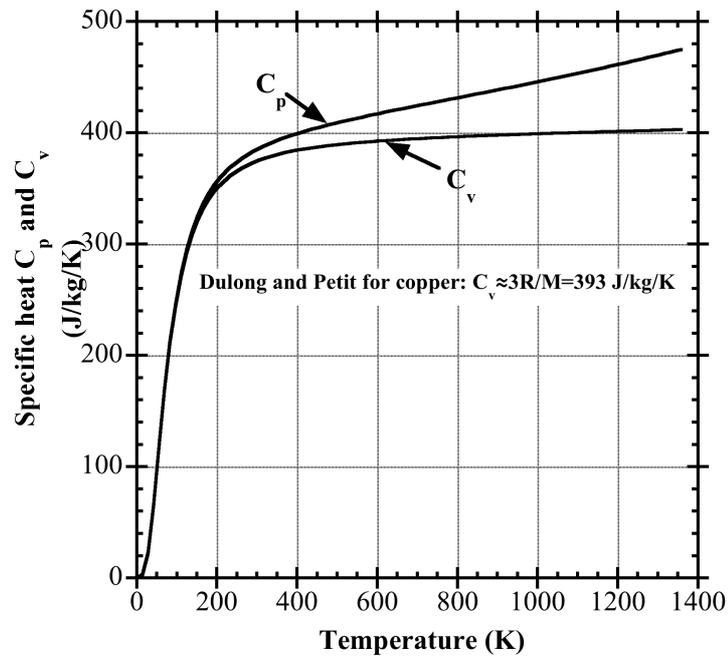

FIG. 1. Specific heat of copper at constant pressure ($C_p$) and at constant volume ($C_v$).

Furthermore, Fig. 1 reveals that between 293 K and the melting point, the specific heat varies linearly with temperature and can be approximated by equation (10):

$$C_P = \frac{3R}{M}[1+\beta(T-293)] = 393.1[1+1.9\times10^{-4}(T-293)]. \qquad (10)$$

For copper, between 77 and 1356K, the resistivity varies linearly with temperature and is classically[25] written as (11), where $\rho_{293K}$ and $\alpha$ are given in Table 1:

$$\rho(T) = \rho_{293K}[1+\alpha(T-293)]. \qquad (11)$$

Matula[30] used the following variant (Eq. 12) of the Bloch-Grünensein equation to calculate the phonons contribution to the resistivity between 4K and the melting temperature :

$$\rho_{ph.}(T) = A\left[1+\frac{BT}{\theta_D - CT}+D(\frac{\theta_D - CT}{T})^p\right]\Phi(\frac{\theta_D - CT}{T}). \qquad (12)$$

With :

$$\Phi(x) = \frac{4}{x^5}\int_0^x \frac{z^5 e^z}{(e^z - 1)^2}dz. \qquad (13)$$

And, for copper[30]:
A=1.809.10$^{-8}$ Ω.m, B=-6.0.10$^{-3}$, C=0.0456, D=-6.476.10$^{-4}$, p=1.84.

According to the Matthiassen rule, impurities and defects (atoms in solid solution, dislocations, vacancies) contribute also to the resistivity through $\rho_i$. Then, the total resistivity can be written as (14):

$$\rho(T) = \rho_{ph.}(T) + \rho_i. \qquad (14)$$

The defects contribution is independent of temperature and led to a plateau at low temperature called Residual Resistivity. The resistivity-temperature curves between 4 and 1356 K copper with different purities are shown in Fig. 2 on a log-log scale which highlights the low temperature region. These data were extracted from[29,30]. The phonon contribution was calculated with equations (12) and (13) to which a residual resistivity of $\rho_i = 2.10^{-11} \Omega.m$ was added[29].

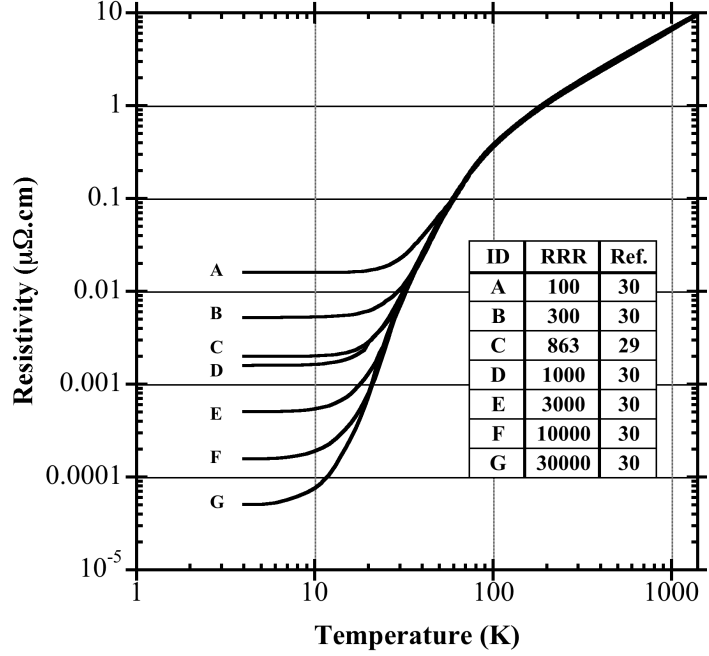

FIG. 2. Resistivity of copper with different purities.

There is no significant difference between the experimental work of Jensen et al.[30], Matula calculations[29], and, for T>293K, the temperature coefficient of resistivity given by Delomel[25]. At low temperature (T<60 K) linearity is no more observed, rather it decreases with a $T^N$ variation with $3<N<5$[28]. Then, resistivity varies according to equation (15):

$$\rho(T) = KT^N + \rho_i . \tag{15}$$

Then, between 4 and 60 K, best fit for the data of Fig. 2 is given by equation (16):

$$\rho(T) \approx 7.88.10^{-17} T^4 + \frac{1.724.10^{-8}}{RRR}. \tag{16}$$

Where RRR is the Residual Resistivity Ratio, i.e. $RRR=\rho_{293K}/\rho_i$. According to the differential form of equation (1), if the current density J is kept constant, the temperature dependence of the heating rate (HR) is given by (17):

$$\frac{dT}{dt} = \frac{J^2 \rho(T)}{D C_p(T)} . \tag{17}$$

The heating rate versus temperature curves are shown Fig. 3, revealing three interesting features. First, it demonstrates that even if J is constant, the heating rate is not constant over the duration of the current pulse when the temperature dependence of the materials properties $C_p$ and $\rho$ are taken into account. Second, the heating rate between 4 and 60 K, the amplitude of the HR variations is several orders of magnitudes and is strongly dependant on the copper purity. Third, in this temperature range, the heating rate decreases to a minimum and then increases up to the melting temperature.

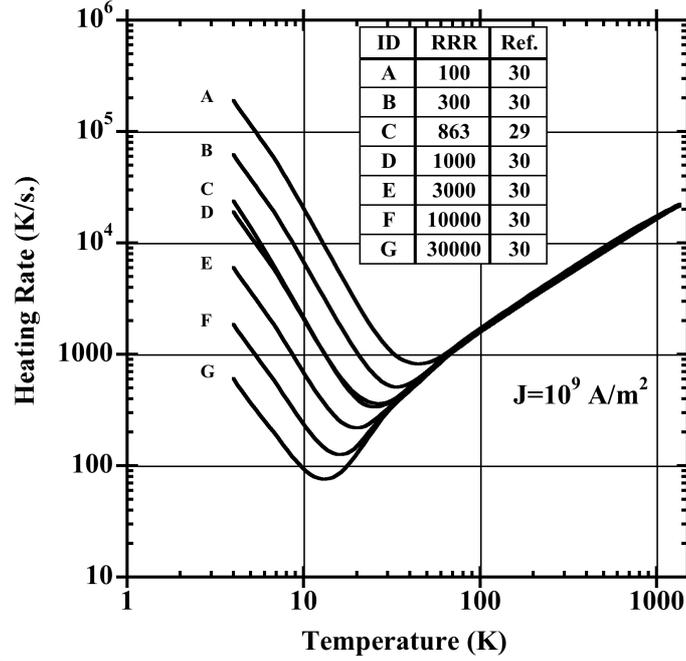

FIG. 3. Heating rate vs. temperature curves of copper during joule Heating. ($J=10^9 A/m^2$).

Equation (17) can be written as (18) which highlights the respective contribution of phonons and impurities to the heating rate:

$$\frac{dT}{dt} = \frac{J^2}{D}\left[\frac{\rho_{ph.}(T)}{C_P(T)} + \frac{\rho_i}{C_P(T)}\right]. \qquad (18)$$

With (19):

$$\left(\frac{dT}{dt}\right)_{ph.} = \frac{J^2}{D}\frac{\rho_{ph.}(T)}{C_P(T)}. \qquad (19)$$

and (20):

$$\left(\frac{dT}{dt}\right)_{imp.} = \frac{J^2}{D}\frac{\rho_i}{C_P(T)}. \tag{20}$$

Fig. 4 reveals the influence of each contribution to the heating rate. Above 77K, the influence of impurities is clearly negligible.

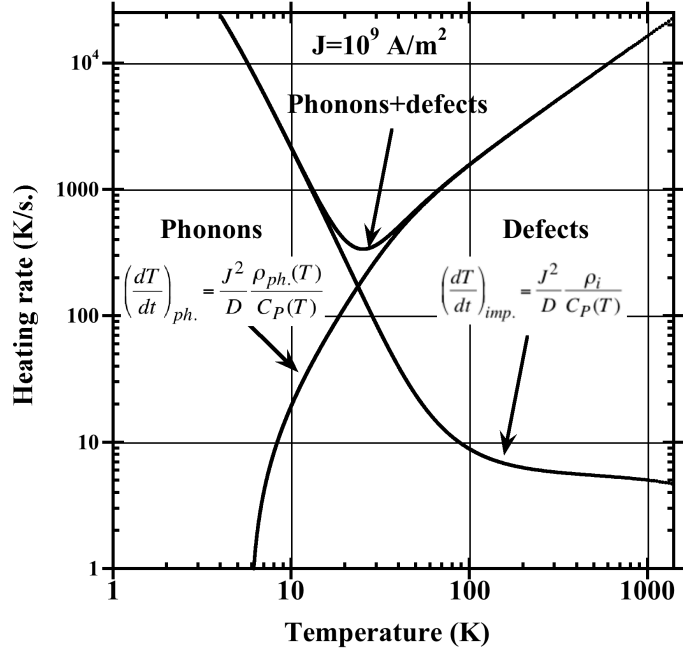

FIG. 4. Heating rate vs. temperature curve of copper during adiabatic Joule heating. This figure hightlights the respective contribution of phonons and impurities ($\rho_i$=0.002 $\mu\Omega$.cm) to the heating rate.

The temperature increase during annealing is calculated by solving equation (21). If J is independent of time it follows :

$$\int_{T_0}^{T_f} \frac{DC_p(T)}{\rho(T)}dT = \int_0^t J^2 dt = J^2 t. \tag{21}$$

The numerical solution of (21) is given in Fig. 5 with three initial temperatures: 4 K, 77 K and 293 K and J=$10^9$ A/m$^2$.

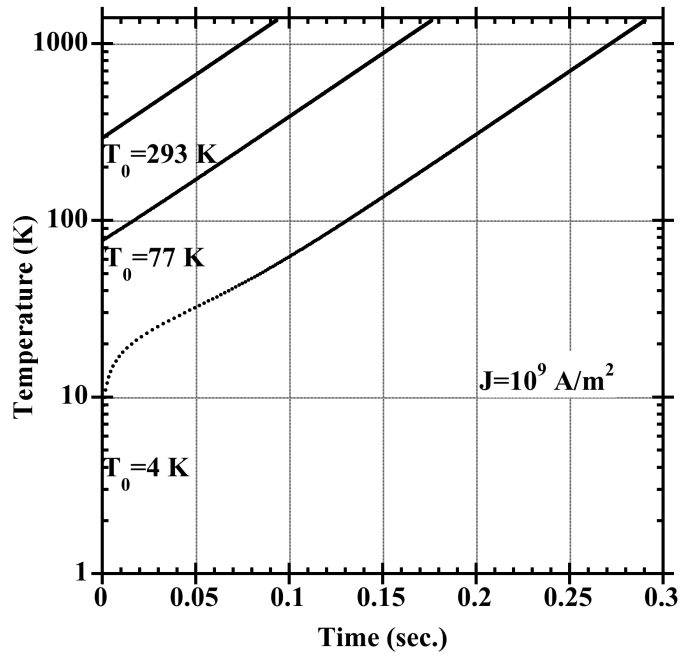

FIG. 5. Temperature vs. time curves of copper during Joule heating, considering three different initial temperatures ($\rho_i$=0.002μΩ.cm).

On a semi-log scale, it is clear that if 77 K<$T_0$<293 K, the temperature vs. time curves are linear. Then, for practical purpose, the following fit can be used to calculate the temperature evolution during Joule heating (22):

$$T \approx T_0 \exp(kt). \tag{22}$$

The best fit for $T_0$=77K and 293K is k=16.28 s$^{-1}$ and 16.47 s$^{-1}$, respectively.

We can search for an analytical solution of (21) for T>293K by considering the temperature variation of $C_p$ and $\rho$. Introducing $C_p(T)$ (eq. 10) and $\rho(T)$ (eq. 11) in the left hand side of equation (21), one obtains (23):

$$\int_{293}^{T} \frac{DC_p(T)}{\rho(T)} dT = \int_{293}^{T} \frac{3RD[1+\beta(T-293)]}{M\rho_{293}[1+\alpha(T-293)]} dT = \frac{3RD}{M\rho_{293}} \int_{293}^{T} \frac{[1+\beta(T-293)]}{[1+\alpha(T-293)]} dT. \tag{23}$$

Setting x=T-293, we integrate (24):

$$\int_{0}^{\Delta T} \left[\frac{1+\beta x}{1+\alpha x}\right] dx = \frac{\beta \Delta T}{\alpha} + \left[\frac{\alpha-\beta}{\alpha^2}\right] \ln(1+\alpha \Delta T). \tag{24}$$

Inserting (24) in (23) and considering the right hand side of equation (21) we obtain (25):

$$\int_{293}^{T} \frac{DC_p(T)}{\rho(T)} dT = \frac{3RD}{\alpha^2 M\rho_{293}} [\alpha\beta\Delta T + (\alpha-\beta)\ln(\alpha\Delta T+1)] = J^2 t. \tag{25}$$

Then setting (26):

$$X = \frac{\alpha^2 M \rho_{293K}}{3R\beta} J^2 t + 1, \tag{26}$$

and (27):

$$Y = \alpha \Delta T + 1, \tag{27}$$

and (28):

$$A = (\frac{\alpha - \beta}{\beta}), \tag{28}$$

we have (29):

$$X = [Y + A\ln(Y)] = \ln(Y^A e^Y) \tag{29}$$

The inverse function of (29) can be found using Mathematica (30):

$$Y = AW\left(\frac{e^{\frac{X}{A}}}{A}\right), \tag{30}$$

Where W(z) is the Lambert function which satisfies, W(z)exp(W(z))=z, this function have several applications in physics[31]. The temperature-time equation can be written as (31):

$$\alpha \Delta T + 1 = \left(\frac{\alpha - \beta}{\beta}\right) W\left\{\left(\frac{\beta}{\alpha - \beta}\right) \exp\left[\left(\frac{\alpha^2 M \rho_{293}}{3RD\beta} J^2 t + 1\right)\left(\frac{\beta}{\alpha - \beta}\right)\right]\right\}, \tag{31}$$

Finally, we have (32):

$$T(K) = 293 + \left(\frac{\alpha - \beta}{\alpha \beta}\right) W\left\{\left(\frac{\beta}{\alpha - \beta}\right) \exp\left[\left(\frac{\alpha^2 M \rho_{293}}{3RD\beta} J^2 t + 1\right)\left(\frac{\beta}{\alpha - \beta}\right)\right]\right\} - \frac{1}{\alpha}. \tag{32}$$

However, if the temperature variation of $C_p$ is neglected (β=0 in eq. (10)), the integration of (21) leads to a much simpler equation (33):

$$T(K) \approx 293 + \frac{1}{\alpha}\left(\exp\left[\frac{\alpha M\rho_{293K}}{3RD}\int_0^t J(t)^2 dt\right] - 1\right). \tag{33}$$

If J is constant during the electrical pulse, (33) can be expressed as (34):

$$T(K) \approx 293 + \frac{1}{\alpha}\left(\exp\left[\frac{\alpha M\rho_{293K}}{3RD}J^2 t\right] - 1\right). \tag{34}$$

The heating rate is therefore (35):

$$\frac{dT}{dt} = \frac{M\rho_{293}}{3RD}\left[\frac{d\left(\int_0^t J(t)^2 dt\right)}{dt}\right]\exp\left[\frac{\alpha M\rho_{293}}{3RD}\int_0^t J(t)^2 dt\right]. \tag{35}$$

Then, knowing the temperature dependence of the materials properties and the current density, the temperature evolution during Joule heating can be calculated. These calculations can be extended to others materials than copper with the exception of ferromagnetic metals (Fe, Ni, Co) which display a strong discontinuity of the specific heat at the Curie temperature[28].

**References**

1   P.H. Frings and L. Van Bocksal, Physica B **211**, 73 (2003).

2   J.D. Embury and K. Hen, Current Opinion in Solid State and Materials Science **3**, 304 (1998).

3   E. Snoeck, F. Lecouturier, L. Thilly, M.J. Casanove, H. Rakoto, G. Coffe, S. Askénazy, J.P. Peyrade, C. Roucau, V. Pantsyrny, A. Shikov and A. Nikulin, Scripta Mat. **38**, 1643 (1998).

4   A. Lagutin, K. Rosseel, F. Herlach, J. Vanacken and Y. Bruynseraede, Meas. Sci. and Technol. **14,** 2144 (2003).

5   A. E. Zielinski, S. Niles and J. D. Powell, J. Appl. Phys. **86**, 3943 (1999).

6   V.I. Oreshkin, S.A. Barengolts and S.A. Chaikovsky in *Proceedings of the 28th International Conference on Phenomena in Ionized Gases*, Prague, Czech Republic, July 15-20, 2007, edited by J. Schmidt, M. Simek, S. Pekarek and V. Prukner (Institute of Plasma Physics, Prague, 2008).

7   V.S. Sedoi and Y.F. Ivanov, Nanotechnology **19**, 145710 (2008).

8   B.Y. Wu, M.O. Alam, Y.C. Chan, J. of Elec. Mat. **37**, 469 (2008).

9   J. Campbell and H. Conrad, Scripta Met. Mat. **31**, 69 (1994).

10  H. Conrad, Mat. Science and Eng. A **287**, 227 (2000).

11  W. Zhang, M.L. Sui, Y.Z. Zhou and D.X. Li, Micron **34**, 189 (2003).

12  Y. Zhou, S. Xiao and J. Guo, Materials Letters **58**, 1948 (2004).

[13] Z. Xu, G. Tang, S. Tian and J. He., Mat. Science and Eng. A **424**, 300 (2006).

[14] H. Lina ,Y. Zhao, B. Zhao, L. Han, J. Ma, Q. Jiang, ISIJ International **48**, 1647 (2008).

[15] Z.J. Wang and H. Song, Trans. Nonferrous Met. Soc. China **19**, s409 (2009).

[16] E.I. Samuel, A. Bhowmik and R.S. Qin, J. Mater. Res. **25**, 1020 (2010).

[17] E. Verdet, *Théorie mécanique de la chaleur, Vol. 2*, (Masson, Paris, 1872), p. 197.

[18] H.S. Carlslaw and J.C. Jaeger, *Conduction of heat in solids*, (Oxford Science Publications, Clarendon Press, Oxford, 1990), p. 149.

[19] M. Nivoit, J.-L. Profizi and D. Paulmier, Int. J. Heat Mass Transfer. **24**, 707 (1981).

[20] V.T. Morgan and N. G. Baron, J. Phys. D : Appl. Phys **19**, 975 (1986).

[21] A.E. Zielinskia, S. Niles and J.D. Powell, Journal of Applied Physics **86**, 3943 (1999).

[22] R. Kratz and P. Wyder, *Principles of pulsed magnet design,* (Springer, Berlin, 2002), p.27.

[23] S. Rembeczki, M. Sc. thesis, University of Debrecen, Debrecen, Hungary (2003).

[24] N. W. Ashcroft and N. D. Mermin, *Physique du Solide,* (Editions de Physique, Les Ulis, 2002), p.537.

[25] J.-C. Delomel, Techniques de l'Ingénieur, vol. K7102v2 (2009).

[26] R.E. Newham, *Properties of Materials Anisotropy, Symmetry, Structure*, (Oxford University Press, Oxford 2005) p. 43.

[27] M. Gerl and J.-P. Issi, *Physique des Matériaux vol. 8*, (Presses Polytechniques et

Universitaires romandes, 1987).

[28] R.E. Smallman and R.J. Bishop, *Modern Physical Metallurgy and Materials Engineering* », (Butterworth-Heinemann, 1999).

[29] R.A. Matula, J. Phys. Chem. Ref. Data **8**, 1147 (1979).

[30] J.E. Jensen, W.A. Tuttle, R.B. Stewart, H. Brechna and A.G. Prodell, Selected cryogenic data notebook, Vol. II, Sections X-XVIII (Brookhaven National Laboratory, Upton, NY 1980), p. X-E-5.

[31] R.M. Corless, G.H. Gonnet, D.E.G. Hare, D.J. Jeffrey and D.E. Knuth, Advances in Computational Mathematics **5**, 329 (1996).